\documentclass[amsmath,amssymb,floatfix,aps,twocolumn,superscriptaddress,nofootinbib,notitlepage,prb]{revtex4-2}

\usepackage{amsmath,amsthm,amssymb,bm,graphicx,float,xcolor,changes,mathtools,enumitem}
\usepackage[colorlinks=true,urlcolor=purple,linkcolor=blue,citecolor=green,bookmarks=false]{hyperref}
\allowdisplaybreaks

\newcommand{\be}{\begin{equation}}
\newcommand{\ee}{\end{equation}}
\newcommand{\6}{\partial}

\newcommand{\la}{\lambda}
\newcommand{\bl}{\boldsymbol{\lambda}}

\newcommand{\bq}{\boldsymbol{q}}
\newcommand{\bmu}{\boldsymbol{\mu}}

\begin{document}

\title{Exact spectral function and nonequilibrium dynamics of the strongly interacting Hubbard model}

\begin{abstract}
Analytical results on the correlation functions of strongly correlated many-body systems are rare in the literature and their importance 
cannot be overstated. We present determinant representations for the space-, time-, and temperature-dependent correlation functions of the 
strongly interacting one-dimensional Hubbard model in the presence of an external trapping potential. These representations are exact and
valid in both equilibrium and nonequilibrium  scenarios like the ones initiated by a sudden change of the confinement potential. In addition,
they can be implemented numerically very easily significantly outperforming other numerical approaches. As applications of our results we 
investigate the single particle spectral functions of systems with harmonic trapping and show that dynamical quasicondensation occurs for 
both fermionic and  bosonic spin-$1/2$ systems released from a  Mott insulator state.
\end{abstract}

\author{Ovidiu I. P\^{a}\c{t}u}
\affiliation{Institute for Space Sciences, Bucharest-M\u{a}gurele, R 077125, Romania}
\author{Andreas Kl\"umper}
\affiliation{Fakult\"at f\"ur Mathematik und Naturwissenschaften, Bergische
 Universit\"at Wuppertal, 42097 Wuppertal, Germany}
\author{Angela Foerster}
\affiliation{Instituto  de  F\'{i}sica  da  UFRGS,  Av.  Bento Gon\c{c}alves  9500,  Porto  Alegre,  RS,  Brazil}

\maketitle

\section{Introduction}

One of the most challenging and still unresolved puzzles in physics is the computation of correlation functions in strongly interacting 
many-body  systems. While significant progress has been achieved in understanding integrable single-component systems in one dimension (1D) 
over the last  decades, both at zero \cite{KBI93,KKMS09,KKMS12} and finite temperature \cite{KBI93,GKS04,GKKK17}, the same level of 
advancement has not been  attained for systems with internal degrees of freedom. The complexity primarily arises from the intricate nature 
of wavefunctions due to the  inclusion of the spin sector, significantly complicating the dynamics.
Moreover, multicomponent systems at low temperatures exhibit diverse additional regimes beyond the conventional Luttinger liquid (LL) phase 
\cite{H81,G03}. These additional phases, such as the spin-incoherent Luttinger liquid (SILL) phase \cite{BL,B1,CZ1,CZ2,Matv,FB,F}, are absent
in their single-component counterparts and have only recently attracted experimental attention \cite{CSKH23}.

In this article, we report on the derivation of determinant representations for the correlation functions of the impenetrable (double occupancy 
is excluded) 1D Hubbard model in the presence of an external potential. Our results are both exact and remarkably versatile, applicable across 
a wide range of conditions, including any temperature, particle statistics, and external potential, whether in equilibrium or non-equilibrium 
scenarios such as those induced by a sudden change in the potential.
Furthermore, our determinant representations are extremely easy to implement numerically, allowing for the investigation of a larger number of 
particles and lattice sites beyond the reach of other numerical techniques like exact diagonalization or time-dependent Density Matrix 
Renormalization Group. The derivation is based on the factorization of the wavefunctions in the strongly interacting regime \cite{OS90,IP98,DFBB08,
GCWM09,VFJV14,LMBP15,YC16,DBBR17,YAP22,MV22} and the connection with a dual system of spinless fermions subjected to the same trapping potential 
or quench (see Appendix~\ref{a1}).

As a first application of our results we have investigated the effect of a harmonic trapping potential on  the single particle spectral 
functions of balanced fermionic and bosonic systems at zero temperature. At zero temperature our results describe the SILL regime and in 
the homogeneous case (no potential) we can highlight the differences between the SILL and LL fermionic spectral functions previously 
investigated in \cite{PHMS96,FHPL97,PHMS97,B23,BP23}. 
As a consequence of non-linear LL theory \cite{IG09a,IG09b} the LL spectral functions exhibit singularities along the dispersion lines of 
the charge and spin excitations \cite{CPMS04,CPSS06,CBP08,E10} a feature which is not present in the SILL spectral function due to the 
spin-incoherence  of the system which smears the singularities. In the fermionic case  the SILL spectral function does not 
exhibit the so-called ``shadow band"  \cite{PHMS96} and the dispersion of the charge part is different compared with the 
LL regime. As a prototypical nonequilibrium scenario 
we study the release of a system from a Mott insulator (MI) state revealing a dynamical quasicondensation phenomenon \cite{RM04,VRSB15} 
characterized by the development of peaks in the momentum distribution at $\pm \pi/(2 a_0)$ ($a_0$ is the lattice constant) as a result of 
emergence of correlations with the same exponents as those characterizing the spin-incoherent system at zero temperature.

\section{The Hubbard model}

In second quantization the Hamiltonian of the  Hubbard model subjected to an external field  is $H=H_K+H_U+H_B+H_E$ \cite{H63,EFGKK}
\begin{subequations}\label{ham}
\begin{align}
H_K=&-t\sum_{z=1}^{L-1}\sum_{\sigma=\{\uparrow,\downarrow\}}\left(a_{z,\sigma}^\dagger a_{z+1,\sigma}+
a_{z+1,\sigma}^\dagger a_{z,\sigma}\right)\, ,\label{hamk} \\
H_U=& U\sum_{z=1}^L n_{z,\uparrow}n_{z,\downarrow}\, , \ \ H_B= B\sum_{z=1}^L(n_{z,\uparrow}-n_{j,\downarrow})\label{hami}\\
H_E=& \sum_{z=1}^L(V(z,t)-\mu)(n_{z,\uparrow}+n_{z,\downarrow})\label{hame}\, ,
\end{align}
\end{subequations}
where $t$ is the hopping strength, $L$ the number of lattice sites, $U>0$ the repulsive Coulomb interaction, $B$ the magnetic field, $\mu$ 
the chemical potential and $n_{z,\sigma}=a_{z,\sigma}^\dagger a_{z,\sigma}$. In the following we will be interested in the case of infinite 
repulsion, $U=\infty$, which means that two particles cannot occupy the same lattice site. We consider the general case of arbitrary statistics 
with $a_{z,\sigma}^\dagger, a_{z,\sigma}$ anyonic fields \cite{SS98,OAU00,CGS16} satisfying  $a_{z,\sigma} a_{m,\sigma'}^\dagger= f(z,m)\,
a_{m,\sigma'}^\dagger a_{z,\sigma}+\delta_{\sigma,\sigma'}\delta_{z,m},$ $a_{z,\sigma}a_{m,\sigma'}=\overline{f}(z,m)a_{m,\sigma'} 
a_{z,\sigma},$ with $f(z,m)=(-1)e^{-i\pi \kappa\, \mbox{\small{sign}}(z-m)}$ (bar denotes complex conjugation)  where $\kappa\in[0,1]$ is the 
statistics parameter and $\mbox{sign}(k)=|k|/k\, ,\mbox{sign}(0)=0$.  As we vary the statistics parameter  $\kappa$ the commutation relations  
interpolate continuously between the anticommutation relations for fermions ($\kappa=0$) and the commutation relations for bosons  ($\kappa=1$). 
We should remark that the statistics parameter can also be chosen  to be $\kappa\in [-1,0]$  with $\kappa=-1$ ($\kappa=0$) describing bosons 
(fermions). At coinciding points the anyonic fields satisfy $(a_{z,\sigma}^\dagger)^2=\left (a_{z,\sigma}\right)^2=0$ (hard-core condition). 
The usual Fermi-Hubbard (Bose-Hubbard) model is obtained for $\kappa=0$ ($\kappa=1$). For the external potentials we will consider either 
static trapping potentials like $V(z)=V_0|z-(L+1)/2|^\alpha$ with $\alpha=1,2,\cdots$  or time dependent potentials of the type $V(z,t=0)=V_I(z)$ 
and $V(z,t>0)=V_F(z)$ which correspond to quantum quenches. In the case of quantum quenches we will denote the initial (final) Hamiltonian by 
$H_I$ ($H_F$). In equilibrium situations we have $H_I=H_F$. We use units of $\hbar=k_B=1$ with $k_B$ the Boltzmann constant.

An important role in our results is played by a dual system of spinless fermions subjected to the same external potential like the Hubbard 
model with the single particle Hamiltonian
\[
H^{SP}=-t\sum_{z=1}^{L-1}\left(|z+1\rangle\langle z|+|z\rangle\langle z+1|\right)+\sum_{z=1}^L V(z,t)|z\rangle\langle z|\, ,
\]
where $|z\rangle=f_z^\dagger|0 \rangle\,$ and $f_z^\dagger,f_m$ are fermionic creation and annihilation operators satisfying $\{f_z,f_m^\dagger\}
=\delta_{z,m}$, $\{f_z,f_m\}=0$. Similar with the case of the Hubbard model in the case of quenches  we will denote the initial (final) single 
particle Hamiltonian  by $H_I^{SP}$ ($H_F^{SP}$). In equilibrium situations at $t=0$ the single particle orbitals satisfy $H_I^{SP}\phi_k(z)= 
\varepsilon(k) \phi_k(z)$ and their time evolution is given by $\phi_k(z,t)=e^{- i\varepsilon(k) t}\phi_k(z)$. In nonequilibrium $\phi_k(z,t)$ is 
the unique time-dependent  solution of the Schr\"odinger equation $i\hbar\6 \phi_k(z,t)/\6t=H_F^{SP}\phi_k(z,t)$ with initial boundary condition 
$\phi_k(z,0)=\phi_k(z)$ [$\phi_k(z)$ is an eigenfunction of $H_I^{SP}$] where $H_F^{SP}$ is the final single particle Hamiltonian.

\section{Correlators} 
We are interested in deriving determinant representations for the space-, time-, and temperature-dependent correlation functions of the Hubbard
model in both equilibrium and nonequilibrium scenarios. The system is initially prepared in a grandcanonical thermal state of the initial 
Hamiltonian  $H_I$  described by the  chemical potential $\mu$, magnetic field $B$ and temperature $T$. The correlators of interest are
\begin{subequations}
\begin{align}
g_\sigma^{(-)}(x,t;y,t')&=
\mbox{Tr} \left[e^{-H_I/T} a_{x,\sigma}^\dagger(t) a_{y,\sigma}(t') \right]/Z\, ,\label{defgm}\\
g_\sigma^{(+)}(x,t;y,t')&=
\mbox{Tr} \left[e^{-H_I/T} a_{x,\sigma}(t) a^\dagger_{y,\sigma}(t') \right]/Z\, ,\label{defgp}
\end{align}
\end{subequations}
where $Z(\mu,B,T) $ is the partition function of the initial thermal state and the time evolved operators are $a_{x,\sigma}(t)=e^{i H_F t} 
a_{x,\sigma}  e^{-i H_F t},$ $a^\dagger_{x,\sigma}(t)=e^{i H_F t} a^\dagger_{x,\sigma} e^{-i H_F t}$. Due to the symmetry   $g^{(\pm)}_\downarrow
(x,t;y,t'|B)=g^{(\pm)}_\uparrow(x,t;y,t'|-B)$ it will be sufficient to study only one type of correlators. The densities and momentum distributions 
are given by  $\rho_\sigma(x,t)=g_\sigma^{(-)}(x,t;x,t)$ and $n_\sigma(k,t)=\sum_{x}\sum_{y} e^{- i k (x-y)} g_\sigma^{(-)}(x,t;y,t)\, .$

\section{Determinant representations}

One of the main results of this article is  the determinant representation for the   correlation functions (for a sketch of the derivation see Appendix~\ref{a2}). 
Introducing the function $F(\gamma,\eta)=1+\sum_{p=1}^\infty\gamma^{-p}\left(e^{i\eta p}+e^{-i\eta p}\right)$ where $\gamma=\left(1+e^{2B/T}\right)$  
the correlators (\ref{defgm}) and (\ref{defgp}) have the representations
\begin{widetext}
\begin{align}\label{detgm}
g_\uparrow^{(-)}(x,t;y,t')&=\frac{e^{-i(t-t')\mu_\uparrow}}{2\pi}\int_{-\pi}^{\pi} F(\gamma,\eta)
\left[\det\left(1+\gamma V^{(T,-)}(\eta)+R^{(T,-)} \right)-\det\left(1+\gamma V^{(T,-)}(\eta) \right)\right]\, d\eta\, ,
\end{align}
\begin{align}\label{detgp}
g_\uparrow^{(+)}(x,t;y,t')&=\frac{e^{i(t-t')\mu_\uparrow}}{2\pi}\int_{-\pi}^{\pi} F(\gamma,\eta)
\left[\det\left(1+\gamma V^{(T,+)}(\eta)-\gamma R^{(T,+)}(\eta) \right)+(g-1)\det\left(1+\gamma V^{(T,+)}(\eta) \right)\right]\, d\eta\, ,
\end{align}
\end{widetext}
where the chemical potentials of the spin-up (-down) particles are $\mu_\uparrow=\mu-B\, (\mu_\downarrow=\mu+B)$ and 
\begin{align}
\left[V^{(T,\pm)}(\eta)\right]_{ab}&=\sqrt{\vartheta(a)}\left(U^{(\pm)}_{ab}(\eta)-\delta_{a,b}\right)\sqrt{\vartheta(b)}\, ,\\
\left[R^{(T,\pm)}\right]_{ab}&=\sqrt{\vartheta(a)}R^{(\pm)}_{ab} \sqrt{\vartheta(b)}\, ,
\end{align}
are infinite matrices.  Here 
\begin{align}
\vartheta(k)=e^{-B/T}/\left(2\cosh(B/T)+e^{(\varepsilon(k)-\mu)/T}\right)\, ,
\end{align}
plays the role of the Fermi function for spin up particles. In Eq.~(\ref{detgp}) $ g\equiv g(x,t;y,t')=\sum_{k=1}^\infty\phi_k(x,t)\overline{\phi}_k(y,t')\, $   and in  terms  of 
the function ($\phi_k(z,t)$ are the time evolved orbitals of the dual system of spinless fermions)
\[
u(k,q|\eta,x,t)=\delta_{k,q}-\left[1-e^{ i (\pi\kappa-\eta)}\right]\sum_{z=x}^\infty\phi_k(z,t)\overline{\phi}_q(z,t)\, ,
\]
the elements of the relevant matrices are given by ($a,b=1,2,\cdots\, $):
\begin{align}
U_{ab}^{(-)}(x,t;y,t'|\,\eta)&=\sum_{q=1}^\infty\overline{u}(a,q|\,\eta,x,t)u(b,q|\,\eta,y,t')\, , \\
U_{ab}^{(+)}(x,t;y,t'|\,\eta)&=\sum_{k=1}^\infty u(k,b|\,\eta,x,t)\overline{u}(k,a|\, \eta,y,t')\, ,
\end{align}
and
\begin{align}
R_{ab}^{(-)}(x,t;y,t')&=\overline{\phi}_a(x,t)\phi_b(y,t')\, ,\\
R_{ab}^{(+)}(x,t;y,t'|\, \eta)&=\overline{e}_a(x,t;y,t'|\, \eta)e_b(x,t;y,t'|\, \eta)\, , 
\end{align}
with 
\begin{align}
e_a(x,t;y,t'|\, \eta)&=\sum_{k=1}^\infty u(k,a|\, \eta,x,t)\overline{\phi}_k(y,t')\, ,\\
\overline{e}_a(x,t;y,t'|\, \eta)&=\sum_{k=1}^\infty\overline{u}(k,a|\, \eta,y,t')\phi_k(x,t).
\end{align}
The representations (\ref{detgm}), (\ref{detgp}) are exact and in practice the numerical computation of the correlators requires only the computation of 
sums, products and determinants of finite size matrices (at zero temperature the matrices have the same dimensions as the number of particles in the system
and at finite temperature one can truncate at level $N_T$ when $\theta(N_T)$ is small enough) which can be performed with extreme efficiency, 
greatly surpassing other numerical schemes 
like exact diagonalization or the time-dependent Density  Matrix  Renormalization Group. For fermionic homogeneous 
systems  with periodic boundary conditions determinant representations were derived in \cite{IPA98,IP98}.

\begin{figure*}[t]
\includegraphics[width=1\linewidth]{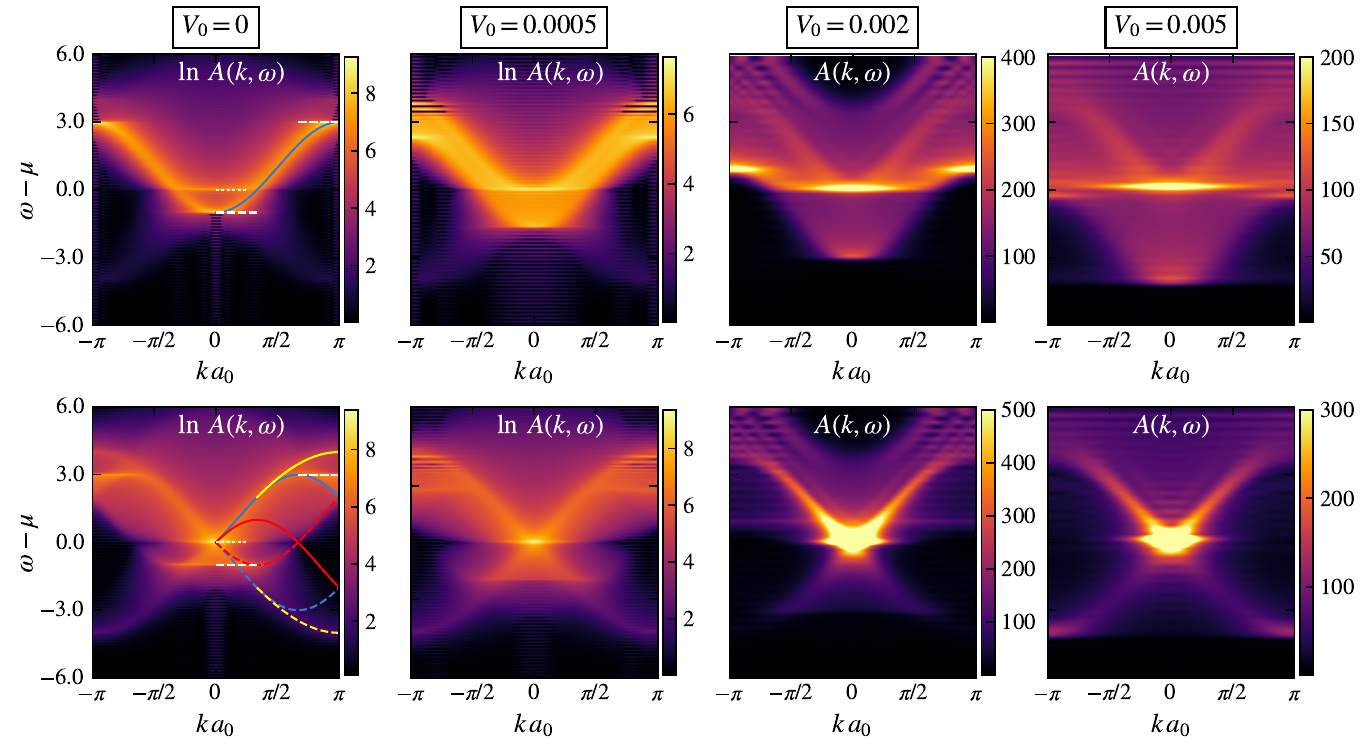}
\caption{Dependence of the spectral function for the zero temperature balanced Hubbard model on the strength of the potential $V_0$ 
($V(z)=V_0|z-(L+1)/2|^2$).  Here $N=40$, $L=120$, $\mu=-1$, and $t=1$. The first row presents results for the fermionic system ($\kappa=0$) and the second row for 
the   bosonic system ($\kappa=1$).  In the first column the continuous blue, red and yellow lines highlight  the $\varepsilon_I(k),$  
$\varepsilon_{II}(k)$  and $\varepsilon_{III}(k)$ singular lines [see Eqs.~(\ref{e1}), (\ref{e2}), (\ref{e3})] while the corresponding dashed lines 
describe $-\varepsilon_I(k),$  $-\varepsilon_{II}(k)$ and $-\varepsilon_{III}(k)$. The dashed white lines are the Van Hove singularities present at 
$\omega=\pm 2 t$  while the dotted white line marks the singularity line present at $\omega=\mu$.
}
\label{fig1}
\end{figure*}

\section{Equilibrium spectral functions}

At equilibrium ($H_I=H_F$) our results can be used to investigate the spectral function of the Hubbard model in the presence of a trapping potential. One 
of the most important quantities in many-body physics, the spectral function encodes information about the accessible energy states and their distribution 
in  momentum space  and can be computed as the imaginary part of the Fourier transform of the retarded Green's function $ A(k,\omega)=-\frac{1}{\pi} 
\mbox{Im}\, G^R(k,\omega), $ with  $G^R(k,\omega)=\int e^{i \omega t}\, dt\sum_x\sum_y e^{- i k(x-y)} G^R(x,t;y,0), $ where the total retarded Green's function is defined
as $ G^{R}(x,t;y,t')=-i \Theta(t-t')\sum_{\sigma\in\{\uparrow,\downarrow\}}\left[g_\sigma^{(+)}(x,t;y,t')+g_\sigma^{(-)}(y,t';x,t)\right].$ 
We will focus on the most interesting case: the balanced system ($B=0$) at zero temperature and investigate the dependence of the spectral function on the 
strength of the harmonic trapping. 
It is important to note that using our representations at zero temperature we investigate the spectral function SILL regime \cite{BL,B1,CZ1,CZ2,Matv,FB,F}  
due to the fact that we take first the limit $U\rightarrow \infty$ and then $T\rightarrow 0$ (for the thermodynamics and quantum criticality in the same regime
see \cite{PKF20,PKF18,LPG23,LPG23b}).
The spectral function in the LL regime \cite{H81,G03} which is 
obtained by taking the limit $T\rightarrow 0$ first and then $U\rightarrow \infty$ has been investigated in \cite{PHMS96,FHPL97,PHMS97,B23,BP23,WWJY24}.

At zero magnetic field and zero temperature the parameter $\gamma=1+e^{2B/T}=2$ and $F(\gamma=2,\eta)\equiv F_0(\eta)=3/(5-4\cos\eta)$. Comparing our results 
with the ones obtained for single component anyons on the lattice \cite{W23,P23} we obtain the following identities ($g_{LL}^{(\pm)}$ are the equivalent 
correlation functions of single component anyons)
\begin{align*}
g^{(-)}_\uparrow(x,t;y,t'|\, \kappa)&=\frac{1}{4\pi}\int_{-\pi}^\pi F_0(\eta)g_{LL}^{(-)}(x,t;y,t'|\, \kappa-\frac{\eta}{\pi})\, d\eta\, , \\
g^{(+)}_\uparrow(x,t;y,t'|\, \kappa)&=\frac{1}{2\pi}\int_{-\pi}^\pi F_0(\eta)g_{LL}^{(+)}(x,t;y,t'|\, \kappa-\frac{\eta}{\pi})\, d\eta\, ,
\end{align*}
which reveal the fact that the correlation functions of the Hubbard model with statistical parameter $\kappa$ can be expressed as integrals over $\eta$ of 
single component anyons of statistics parameter $\kappa-\eta/\pi$ with weight $\frac{1}{2\pi}\frac{3}{5-4\cos\eta}$. In the fermionic case similar identities 
were derived using a different method in \cite{GQBZ23}. 
The previous relations also show that the spectral function of the Hubbard model in the SILL regime is closely related to the spectral function of single 
component anyons. We remind the reader that  the singular lines of the spectral function for single component anyons are given by \cite{SGPM21,W23,P23}
\begin{subequations}
\begin{align}
\varepsilon_{I}(k)&=-2t\cos(k+k_F\kappa)-\mu\, ,\label{e1}\\
\varepsilon_{II}(k)&=2t\cos(k+k_F\kappa-2k_F)+\mu\, ,\label{e2}\\
\varepsilon_{III}(k)&=4t\sin[(k+k_F\kappa-k_F)/2]\, ,\label{e3}
\end{align}
\end{subequations}
where $k_F=\pi N/L$ is the Fermi vector with $N$ the number of particles. In the case of single component anyons decreasing the statistics parameter from 
$\kappa=1$  (bosons) the spectral weight from  $\varepsilon_{II}(k)$ and $\varepsilon_{III}(k)$ also  decreases being transferred to $\varepsilon_{I}(k)$ 
which becomes the only singular line at $\kappa=0$  as it is expected for a free fermionic system.

In Fig.~\ref{fig1} we present results for the spectral function of the Hubbard model (fermionic and bosonic) in the presence of a harmonic potential 
$V(z)=V_0|z-(L+1)/2|^2$  with $V_0$ varying from zero (homogeneous system) to $V_0 = 0.005$ a value for which the system is in the Mott insulator phase 
(the middle of the trap contains a region in which the density is 1). While we are not aware of any results reported in the literature for the spectral function of the 
Hubbard model subjected to an external trapping potential in the homogeneous case the spectral function for the Fermi-Hubbard model in the LL regime has been 
investigated in  \cite{PHMS96,FHPL97,PHMS97,B23,BP23,WWJY24}.  This gives us the opportunity to highlight the differences between the spectral 
functions in the two regimes. 
Both spectral functions  present Van Hove singularities at $\pm 2t$ (in the vicinities of $k=\pm \pi$ and $k=0$) and an additional singularity at 
$\omega=\mu$ (around $k=0$). However, the SILL spectral function  does not exhibit the so-called  ``shadow band" \cite{PHMS96} and the dispersion of the charge 
part for $A(k,\omega)$ is given by Eq.~(\ref{e1}) compared with $\varepsilon_{LL}(k)=-2 t\cos(|k|+k_F^{LL}), (k_F^{LL}= k_F/2)$ for the LL spectral function. 
Another major difference is that while in the LL regime the  spectral function has a power-law behaviour  \cite{CPMS04,CPSS06,CBP08,E10} $A(k,\omega)\sim|\omega-
\varepsilon_{j}(k)|^{-\mu_j}$  near each singular line $\varepsilon_j(k)$ with exponent $\mu_j$ that is momentum dependent  this feature is absent in the SILL phase. 
This is due to the 
fact that $A(k,\omega)$ in the SILL phase can be expressed as an integral over the spectral functions of single component anyonic systems which smoothens
the ``singular lines"  in both  fermionic and bosonic cases [Eqs.~(\ref{e1}), (\ref{e2})].
The presence of the  trapping potential significantly changes the characteristics of the spectral function. For small values of the curvature ($V_0=0.0005$) 
the spectral function  presents similar features like in the homogeneous case but with an increased $k_F$ coupled with a broadening of  the pseudo-singular 
lines [Eqs.~(\ref{e1}),  (\ref{e2}) and (\ref{e3})]. However, for stronger potentials the homogeneous picture washes out with the majority of the spectral weight 
migrating along the $\omega=\mu$ singular line and the origin. 

\begin{figure}[h]
\includegraphics[width=1\linewidth]{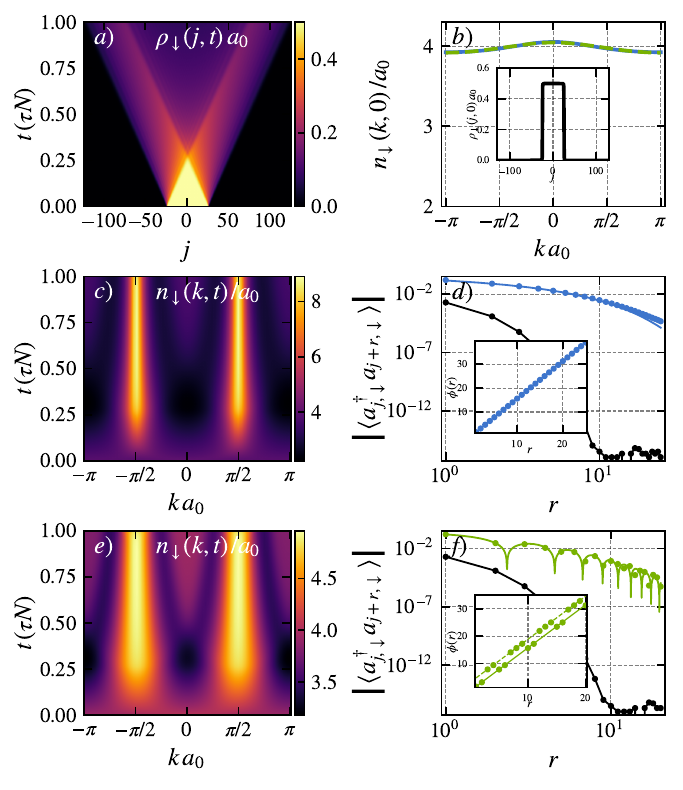}
\caption{Expansion of the zero temperature balanced Hubbard model from a Mott insulator initial state. Here $N=50$, $L=250$ and $\tau=\hbar/t$. 
a) Time evolution of the density $\rho_\downarrow(j,t)=\rho_\uparrow(j,t)$, which is independent of statistics, after the trapping potential is removed.
b) Initial momentum densities $n_\downarrow(k,0)= n_\uparrow(k,0)$ for  fermions (green dashed line) and bosons (blue continuous line). Inset: Density 
profile 
$\rho_\uparrow(j,0)=\rho_\downarrow(j,0)$ at $t=0$ which shows that the system is in the MI state in the middle of the trap.
c) Time evolution after expansion of the momentum distribution $n_\downarrow(k,t)$ for a bosonic system $\kappa=1$ from which we can see clearly the peaks at 
$\pm \pi/(2 a_0)$. 
d) $\left|\langle a_{j,\uparrow}^\dagger a_{j+r,\uparrow}\rangle\right|$  for $j=26$ at $t=0$ (black markers, the black continuous line is a guide to the eye) 
and  $t=0.4 N\tau$ (blue markers) with the asymptotic  value (continuous blue  line) $\mathcal{A}(r)$ given by Eq.~(\ref{asymbos}) with $a=0.20080, 
k_F^*=1.48750$.  
Inset: The phase pattern of the bosonic correlator  (blue markers), the blue continuous line is $\frac{\pi}{2} r$. 
e) Time evolution after expansion of the momentum distribution $n_\downarrow(k,t)$ for a fermionic system $\kappa=0$. 
f)  $\left|\langle a_{j,\uparrow}^\dagger a_{j+r,\uparrow}\rangle\right|$  for $j=26$  at $t=0$ (black markers, the black continuous line is a guide to the eye) 
and  $t=0.4 N\tau$ (blue markers) and the asymptotic value (continuous blue line)  $\mathcal{A}(r)$ given by Eq.~(\ref{asymfer}) with $a=0.22646, k_F^*=1.48228,  
\phi=0.14982$. 
Inset: The phase pattern of the fermionic correlator (green markers), the green continuous lines are $\frac{\pi}{2} r$ and $\frac{\pi}{2} r+\pi$. 
}
\label{fig2}
\end{figure}

\section{Dynamical quasicondensation}

Our determinant representations for the correlators (Eqs.~\ref{detgm} and \ref{detgp}) remain valid in experimentally accessible nonequilibrium scenarios, such as those 
initiated by a sudden change in the trapping potential. In nonequilibrium situations, like the release from a trap or the evolution from a domain wall boundary, it is 
sufficient to focus on the equal-time correlation function, $g_\sigma^{(-)}(x,t;y,t')$ at $t=t'$. From this function, we can extract the dynamics of real-space densities 
and momentum distributions.
At equal times,  $t=t'$, the integration over the variable $\eta$ can be performed analytically, yielding the following determinant representation 
[$g^{(-)}_\uparrow(x,y|\, t)\equiv g^{(-)}_\uparrow(x,t;y,t)$]
\be\label{etmain}
g^{(-)}(x,y|\, t)=\det\left(1+v^{(T,-)}+r^{(T,-)}\right)-\det\left(1+v^{(T,-)}\right)\, ,
\ee
with the matrices $v^{(T,-)}$ and $r^{(T,-)}$  given by    
\begin{subequations}\label{etdet}
\begin{align}
v^{(T,-)}_{ab}&=-\left[\gamma-e^{-i \pi\kappa\,\mbox{\small{sign}}(y-x)}\right]\sqrt{\vartheta(a)\vartheta(b)}\nonumber\\
&\qquad\qquad\qquad\times 
\sum_{z=\min(x,y)}^{\max(x,y)-1}\overline{\phi}_a(z,t)\phi_b(z,t)\, ,\\\
r^{(T,-)}_{ab}&=\sqrt{\vartheta(a)\vartheta(b)}\,\,\overline{\phi}_a(x,t)\phi_b(y,t)\, .
\end{align}
\end{subequations}
where $a,b=1,2,\cdots$.
It is important to note that $g^{(-)}(x,y|\, t)=\overline{g^{(-)}(y,x|\, t)}$. 
At zero temperature and zero magnetic field (at $T=0$ even an infinitesimal magnetic field   completely polarizes the system, making it equivalent to a single-component system) 
we have $\gamma=2$ and $\vartheta(q)=1/2$ if $\varepsilon(q)\le \mu$ and zero otherwise. In this case, the relevant matrices are $N$-dimensional where $N$
represents the number of energy levels with energies smaller than the chemical potential. The matrix elements for $a,b=1,\cdots, N$ are given by
\begin{align}
v^{(0,-)}_{ab}&=-\frac{1}{2}\left[\gamma-e^{-i \pi\kappa\,\mbox{\small{sign}}(y-x)}\right]\nonumber\\
&\qquad\qquad\qquad\times
\sum_{z=\min(x,y)}^{\max(x,y)-1}\overline{\phi}_a(z,t)\phi_b(z,t)\, ,\\
r^{(0,-)}_{ab}&=\frac{1}{2}\,\,\overline{\phi}_a(x,t)\phi_b(y,t)\, .
\end{align}

One nonequilibrium scenario of considerable interest is represented by the free expansion of a Mott insulator 
 state located initially in the center of the trap (for other nonequilibrium scenarios see  \cite{IN17,BTC17,RBC22,TCB23,GHPZ23}). In Fig.~\ref{fig2} we present 
the time evolution of the  densities and momentum distributions for a system of $N=50$ particles on a lattice with $L=250$ sites after removal of the harmonic 
trapping potential with the system initially prepared in a MI state. While the initial quasimomentum distribution is almost flat (for an infinite lattice the 
distribution is flat) and there are no off-diagonal correlations $\langle a_{j,\sigma}^\dagger a_{j+r,\sigma}\rangle\sim 0$, after release from the trap the 
momentum distributions develop  peaks at $\pm \pi/(2 a_0)$ which are more pronounced in the bosonic case [see Fig.~\ref{fig2}c) for the bosonic case 
and Fig.~\ref{fig2}e) for the fermionic case]. This rather 
counterintuitive phenomenon is the multi-component counterpart of the dynamical quasicondensation of hard-core bosons which was theoretically predicted in \cite{RM04} 
(see also \cite{HRMF08,VIR17,TSCV24,BSDR24}) and validated experimentally in \cite{VRSB15}.  In the case of single component bosons  it was argued \cite{VRSB15} that these 
singularities are due to emerging power-law correlations $\langle a_j^\dagger(t) a_{j+r}(t)\rangle=\mathcal{A}(r)e^{i \Phi(r)},$ $A(r)\sim r^{-1/2},$ $\Phi(r)=
\pm \frac{\pi}{2} r,$ with the plus (minus) sign for the right (left) expanding cloud. It is interesting to note that even though we are in a nonequilibrium situation 
the power law decay of the absolute value is given by the same exponent as in the groundstate ($1/2$) but with an alternating phase pattern $\pm \pi/2$ between 
neighbouring sites producing peaks at finite quasimomenta (for each half system).
Guided by the single component case in which the emerging correlation functions have the same asymptotic behavior like in the groundstate with an alternating phase 
we have discovered that the same phenomenon remains valid in the two-component case. In the bosonic case we have  $\langle a_{j,\uparrow}^\dagger(t)  a_{j+r,\uparrow}
(t)\rangle=\mathcal{A}(r)e^{i \Phi(r)}$ with the asymptotic behavior [see Fig.~\ref{fig2}d)]
\be\label{asymbos}
\mathcal{A}(r)\sim a\, e^{-\nu k_F^* r}r^{-\frac{1}{2}+\frac{1}{2}\nu^2}\, ,\ \ \Phi(r)=\pm \frac{\pi}{2} r\, ,
\ee
where $a$ and $k_F^*$ are free parameters [$k_F^*$ is very likely related to the density of the particle cloud at $(j,t)$] and $\nu=\frac{\ln 2}{\pi}$ which has the same 
form as the large distance asymptotics of two-component bosons in the SILL 
regime  \cite{CSZ05,P19}. It is important to note that because we are in the spin-incoherent regime the  asymptotics are exponentially decreasing even at zero temperature 
with an  exponent depending on the number of components which is a general feature of incoherent systems \cite{F}. However, 
the power exponent ($\frac{1}{2}\nu^2 \sim 0.024$) is close to 0, the value for single component bosons
and even though the momentum distribution does not present singularities at $\pm \pi/(2 a_0)$ we still have a 
sizeable number of particles with momenta concentrated around $\pm \pi/(2 a_0)$.
In the fermionic case the asymptotics are \cite{CZ08} [see  Fig.~\ref{fig2}f)]
\be\label{asymfer}
\mathcal{A}(r)\sim a\, e^{-\nu k_F^* r}r^{-1+\frac{1}{2}\nu^2}|\sin\left(k_F^* r-\nu \ln r-\phi \right)|\, ,
\ee
with $a, k_F^*$ and $\phi$ free parameters again mimicking the results for two-component fermions at zero temperature in the spin-incoherent regime. In this case the 
phase  is $\Phi(r)=\pm \frac{\pi}{2} r$ plus a contribution of $\pi$ whenever the $\sin$ factor in (\ref{asymfer}) is negative.

\section{Conclusions}
We have introduced versatile determinant representations for the correlation functions of the strongly interacting Hubbard model, which facilitated the numerical investigation 
of the spectral function in the presence of a confinement potential and the dynamics of the system after release from the trap. Our results are highly general, easy to implement 
numerically, and can be used to study the static properties in different trapping geometries, the nonequilibrium dynamics in various scenarios (periodic modulation of the trapping 
frequency, quenches of the confining potential, etc.), the presence or absence of dynamical localization in periodically kicked systems, and the transport properties at zero and 
finite temperature. 
Additionally, it is worth noting that our equal-time  representation, Eq.~\ref{etmain}, bears a strong resemblance to Lenard's formula \cite{L66}, which allows for rigorous analytic 
results concerning asymptotic behavior to be derived.

\acknowledgments

O.I.P. acknowledges financial support from the Grant No. 30N/2023 of the National Core Program of the Romanian Ministry of Research, Innovation and Digitization.
A.K. acknowledges financial support by Deutsche Forschungsgemeinschaft through FOR 2316. The writing of this paper was carried out while A.K. and A.F.
participated in the MATRIX workshop MPI2024. A.K. also gratefully acknowledges support through the PIFI fellowship by the Chinese Academy of Sciences and the
Innovation Academy for Precision Measurement Science and Technology, Wuhan, where the final writing was completed.
A.F. acknowledges CNPq (Conselho Nacional de Desenvolvimento Cient\'{i}fico e Tecnol\'{o}gico) for financial support.

\appendix

\begin{widetext}

\section{Eigenstates at $t=0$ and their time evolution}\label{a1}

In the impenetrable limit the doubly occupied states have infinite energy and are effectively excluded from the spectrum. In this limit the dynamics 
of the system is described by the effective Hamiltonian
\begin{align}\label{hamt}
\tilde{H}= P\left[-t\sum_{z=1}^{L-1}\sum_{\sigma=\{\uparrow,\downarrow\}}\left(a_{z,\sigma}^\dagger a_{z+1,\sigma}+a_{z+1,\sigma}^\dagger a_{z,\sigma}
\right)+B\sum_{z=1}^L(n_{z,\uparrow}-n_{z,\downarrow})+ \sum_{z=1}^L(V(z,t)-\mu)(n_{z,\uparrow}+n_{z,\downarrow})\right]P\, ,
\end{align}
with
\be
P=\prod_{z=1}^L\left(1-n_{z,\uparrow}n_{z,\downarrow}\right)\, , \ \ P^\dagger=P\, ,\ \ P^2=P\, ,
\ee
the projector on singly occupied states. In the impenetrable regime the physical Hilbert space denoted by $\mathfrak{H}$ has dimension  $3^L$ while the 
space $\mathfrak{F}$ in which the canonical Hubbard operators act has dimension $4^L$. In the time dependent case we will  denote the initial (final) 
Hamiltonian by $\tilde{H}_I$ ($\tilde{H}_F$).

At $t=0$ the eigenstates of the initial Hamiltonian (\ref{hamt}) in the $(N,M)$-sector ($N$ particles of which $M$ have spin down) have the form
\begin{align}\label{eigen}
|\boldsymbol{\psi}_{N,M}(\boldsymbol{k},\boldsymbol{\lambda})\rangle=\sum_{z_1,\cdots,z_N=1}^L\sum_{\alpha_1,\cdots,\alpha_N=\{\uparrow,\downarrow\}}^{[N,M]}\,
\chi_{N,M}^{\boldsymbol{\alpha}}(z_1,\cdots,z_N|\boldsymbol{k},\boldsymbol{\lambda})a_{z_N,\alpha_N}^\dagger \cdots a_{z_1,\alpha_1}^\dagger|0\rangle\, ,
\end{align}
where $[N,M]$ means that the sum is over all combinations of $N$ $\alpha$'s of which $M$ are equal to $\downarrow$ and $N-M$ are $\uparrow$,
$\boldsymbol{\alpha}=(\alpha_1,\cdots, \alpha_N)$ and  $|0\rangle$ is the Fock vacuum satifying $a_{z,\alpha} |0\rangle=\langle 0|a_{z,\alpha}^\dagger=0$
for all values of $z$ and $\alpha$ and $\langle 0|0\rangle=1$.The eigenstates (\ref{eigen}) are parameterized by two sets of unequal (in each set) numbers
$\boldsymbol{k}=\{k_j\}_{j=1}^N$ and $\boldsymbol{\lambda}=\{\lambda_j\}_{j=1}^M$ which characterize the charge and spin degrees of freedom.
The total wavefunctions are 
\begin{align}\label{wavef}
\chi_{N,M}^{\boldsymbol{\alpha}}(\boldsymbol{z}|\boldsymbol{k},\boldsymbol{\lambda})=\frac{1}{N!}
\left[\sum_{P\in S_N} \xi_{XX}^{(\boldsymbol{\alpha}, P\boldsymbol{\alpha})} (\boldsymbol{\lambda})
e^{i\frac{\pi\kappa}{2}\sum_{1\le a<b\le N}\epsilon(z_a-z_b)}\theta(z_{P(1)}<\cdots <z_{P(N)})\right]\det_N\left[\phi_{k_a}(z_b)\right]\, ,
\end{align}
with  $\theta(z_{P(1)}<\cdots <z_{P(N)})$  equal to $1$ when $z_{P(1)}<\cdots <z_{P(N)}$ and $0$ otherwise and $\epsilon(z)=1$ for $z\ge 0$ and $\epsilon(z)=-1$ 
for $z<0$. The spin sector is described by  $\xi_{XX}^{(\boldsymbol{\alpha}, P\boldsymbol{\alpha})} (\boldsymbol{\lambda})$, the wavefunctions of the XX spin chain 
with periodic boundary conditions, defined by 
\begin{align}\label{wavefxx}
\xi_{XX}^{(\boldsymbol{\alpha}, P\boldsymbol{\alpha})} (\boldsymbol{\lambda})=
\frac{1}{N^{M/2}}\left[\prod_{1\le j<l\le M} \mbox{sign}(n_l-n_j)\det_M\left(e^{i \lambda_a n_b}\right)\right]\, ,
\end{align}
with the Bethe ansatz equations 
\begin{align}\label{baexxs}
e^{i\lambda_l N}&=(-1)^{M+1}\, ,\ \ l=1,\cdots,M\, .\
\end{align}
In (\ref{wavefxx})  $n_l$ is the position of the $l$-th spin down on the auxiliary lattice of the XX spin chain. For example, in the $(N,M)=(4,2)$ sector and
$\boldsymbol{\alpha}=(\alpha_1,\alpha_2,\alpha_3,\alpha_4)=(\downarrow\downarrow\uparrow\uparrow)$ for the permutation $P=(1243)$ we have $P\boldsymbol{\alpha}
=(\alpha_{P(1)},\alpha_{P(2)},\alpha_{P(3)},\alpha_{P(4)})=(\downarrow\downarrow\uparrow\uparrow)$ and $(n_1,n_2)=(1,2)$ while for $P=(3214)$ we have 
$P\boldsymbol{\alpha}=(\uparrow\downarrow\downarrow\uparrow)$ and $(n_1,n_2)=(2,3)$.
The charge sector is described by a Slater determinant of eigenfunctions of the  single particle Hamiltonian with open boundary conditions ($\{f_i,f_j^\dagger\}=\delta_{i,j}\, ,$ 
$\{f_i,f_j\}=0\, ,$ are spinless fermions)
\be\label{hamspi1}
H_I^{SP}=-t\sum_{z=1}^{L-1}\left(|z+1\rangle\langle z|+|z\rangle\langle z+1|\right)+\sum_{z=1}^L V(z,t\le 0)|z\rangle\langle z|\, ,
\ \ \mbox{ with } |z\rangle=f_z^\dagger|0 \rangle\, .
\ee
The single particle wavefunctions satisfy $H_I^{SP}\phi_k(z)=\varepsilon(k)\phi_k(z)$ and  we have 
\be
\tilde{H}_I|\boldsymbol{\psi}_{N,M}(\boldsymbol{k},\boldsymbol{\lambda})\rangle=E_{N,M}(\boldsymbol{k})
|\boldsymbol{\psi}_{N,M}(\boldsymbol{k},\boldsymbol{\lambda})\rangle\, ,\  \  \   E_{N,M}(\boldsymbol{k})=
\sum_{j=1}^N\varepsilon(k)-\mu N+B(2N-M)\, .
\ee
The wavefunctions (\ref{wavef}) are the natural generalization of the Bethe ansatz wavefunctions \cite{IPA98,IP98}, satisfy the many-body Schr\"odinger equation with 
open boundary conditions, have the appropriate symmetry and they form a complete orthonormal system of dimension $3^L$.

The time evolved eigenstates are also described by (\ref{eigen}) with the time-dependent wavefunctions
\begin{align}\label{waveft}
\chi_{N,M}^{\boldsymbol{\alpha}}(\boldsymbol{z},t|\boldsymbol{k},\boldsymbol{\lambda})=\frac{1}{N!}
\left[\sum_{P\in S_N} \xi_{XX}^{(\boldsymbol{\alpha}, P\boldsymbol{\alpha})} (\boldsymbol{\lambda})
e^{i\frac{\pi\kappa}{2}\sum_{1\le a<b\le N}\epsilon(z_a-z_b)}\theta(z_{P(1)}<\cdots <z_{P(N)})\right]\det_N\left[\phi_{k_a}(z_b,t)\right]\, ,
\end{align}
where $\phi_k(z,t)=e^{- i\varepsilon(k) t}\phi_k(z)$ in equilibrium situations. In nonequilibrium situations $\phi_k(z,t)$ is the unique time-dependent 
solution of the Schr\"odinger equation $i\hbar\6 \phi_k(z,t)/\6 t=H_F^{SP}\phi_k(z,t)$ with initial boundary condition $\phi_k(z,0)=\phi_k(z)$ [remember 
$\phi_k(z)$ is an eigenfunction of (\ref{hamspi1})] where $H_F^{SP}$ is the final single particle Hamiltonian
\be
H_F^{SP}=-t\sum_{z=1}^{L-1}\left(|z+1\rangle\langle z|+|z\rangle\langle z+1|\right)+\sum_{z=1}^L V(z,t> 0)|z\rangle\langle z|\, ,\ \ \mbox{ with } 
|z\rangle=f_z^\dagger|0 \rangle\, .
\ee

\section{Sketch of the derivation for the determinant representations}\label{a2}

At finite temperature the correlation functions of the system prepared in a thermal state of the initial Hamiltonian $\tilde{H}_I$ described by 
the grandcanonical potential at  chemical potential $\mu$, magnetic field $B$, and temperature $T$ are
\begin{align}
g_\sigma^{(-)}(x,t;y,t')&=\langle \tilde{a}_{x,\sigma}^\dagger(t) \tilde{a}_{y,\sigma}(t')\rangle_{\mu,B,T}\, ,\ \ \ \sigma=\{\uparrow,\downarrow\}\, ,\nonumber\\
&=\mbox{Tr} \left[e^{-\tilde{H}_I/T} \tilde{a}_{x,\sigma}^\dagger(t) \tilde{a}_{y,\sigma}(t') \right]/\mbox{Tr} \left[e^{-\tilde{H}_I/T} \right]\, ,\nonumber\\
&=\sum_{N=0}^\infty \sum_{M=0}^{N+1} \sum_{\substack{k_1<\cdots<k_{N+1}\\  \la_1<\cdots<\la_{M}}} \frac{e^{-E_{N+1,M}(\boldsymbol{k})/T}}{\mathcal{Z}(\mu,B,T)}
\langle\boldsymbol{\psi}_{N+1,M}(\boldsymbol{k},\boldsymbol{\la})|\tilde{a}_{x,\sigma}^\dagger(t) \tilde{a}_{y,\sigma}(t')|\boldsymbol{\psi}_{N+1,M}(\boldsymbol{k},
\boldsymbol{\la})\rangle\, ,\label{defgma}\\
g_\sigma^{(+)}(x,t;y,t')&=\langle \tilde{a}_{x,\sigma}(t) \tilde{a}^\dagger_{y,\sigma}(t')\rangle_{\mu,B,T}\, ,\ \ \ \sigma=\{\uparrow,\downarrow\}\, ,\nonumber\\
&=\mbox{Tr} \left[e^{-\tilde{H}_I/T} \tilde{a}_{x,\sigma}(t) \tilde{a}^\dagger_{y,\sigma}(t') \right]/\mbox{Tr} \left[e^{-\tilde{H}_I/T} \right]\, ,\nonumber\\
&=\sum_{N=0}^\infty \sum_{M=0}^{N} \sum_{\substack{q_1<\cdots<q_{N}\\  \mu_1<\cdots<\mu_{M}}} \frac{e^{-E_{N,M}(\boldsymbol{q})/T}}{\mathcal{Z}(\mu,B,T)}
\langle\boldsymbol{\psi}_{N,M}(\boldsymbol{q},\boldsymbol{\mu})|\tilde{a}_{x,\sigma}(t) \tilde{a}^\dagger_{y,\sigma}(t')|\boldsymbol{\psi}_{N,M}(\boldsymbol{q},
\boldsymbol{\mu})\rangle\, .\label{defgpa}
\end{align}
where $\mathcal{Z}(\mu,B,T)=\sum_{N=0}^\infty \sum_{M=0}^{N} \sum_{\substack{q_1<\cdots<q_{N}\\  \mu_1<\cdots<\mu_{M}}} e^{-E_{N,M}(\boldsymbol{q})/T}=
\prod_{q=1}^\infty \left(1+2\cosh(B/T)e^{-(\varepsilon(q)-\mu)/T}\right) $ is
the partition function of the initial thermal state and the time evolved operators are
\be
\tilde{a}_{x,\sigma}(t)=e^{i \tilde{H}_F t} \tilde{a}_{x,\sigma} e^{-i \tilde{H}_F t}\, , \ \
\tilde{a}^\dagger_{x,\sigma}(t)=e^{i \tilde{H}_F t} \tilde{a}^\dagger_{x,\sigma} e^{-i \tilde{H}_F t}\, , \ \
\ee
with
\be
\tilde{a}_{x,\sigma}=P a_{x,\sigma} P\, ,\ \ \tilde{a}^\dagger_{x,\sigma}=P a^\dagger_{x,\sigma} P\, .
\ee
It is important to note that the traces in (\ref{defgma}) and (\ref{defgpa}) are taken in the Hilbert space $\mathfrak{H}$ of dimension $3^L$.

The derivation of the determinant representations, Eqs.~\ref{detgm} and \ref{detgp}, is based on the summation of the form factors \cite{IPA98,IP98}. Using a resolution of the 
identity, $ \bm{1}=\sum_{N=0}^\infty\sum_{M=0}^N \sum_{\substack{q_1<\cdots<q_N\\  \mu_1<\cdots<\mu_M}}|\Phi_{N,M}(\bq,\bmu)\rangle\langle\Phi_{N,M} (\bq,\bmu)|\, ,$ the mean 
values of bilocal operators appearing in the right hand-side of (\ref{defgm}) and (\ref{defgp}) can be  written as 
\begin{align}\label{i30}
\langle \boldsymbol{\psi}_{N+1,M}(\boldsymbol{k},\bl)|\tilde{a}_{x,\sigma}^\dagger(t) \tilde{a}_{y,\sigma}(t')|\boldsymbol{\psi}_{N+1,M}(\boldsymbol{k},\bl)\rangle
&=\sum_{\substack{q_1<\cdots<q_N\\  \mu_1<\cdots<\mu_{\bar{M}}}}
\overline{\mathcal{F}}_{N,M}^{(\sigma)}(\boldsymbol{k},\bl;\bq,\bmu|x,t)
\mathcal{F}_{N,M}^{(\sigma)}(\boldsymbol{k},\bl;\bq,\bmu|y,t')\, ,
\end{align}
and
\begin{align}\label{i31}
\langle \boldsymbol{\psi}_{N,\bar{M}}(\bq,\bmu)|\tilde{a}_{x,\sigma}(t) \tilde{a}^\dagger_{y,\sigma}(t')|\boldsymbol{\psi}_{N,\bar{M}}(\bq,\bmu)\rangle
&=\sum_{\substack{k_1<\cdots<k_{N+1}\\  \lambda_1<\cdots<\lambda_M}}
\mathcal{F}_{N,M}^{(\sigma)}(\boldsymbol{k},\bl;\bq,\bmu|x,t)
\overline{\mathcal{F}}_{N,M}^{(\sigma)}(\boldsymbol{k},\bl;\bq,\bmu|y,t')\, ,
\end{align}
where
\be\label{formf}
\mathcal{F}_{N,M}^{(\sigma)}(\boldsymbol{k},\bl;\bq,\bmu|x,t)=\langle\Psi_{N,\bar{M}}(\bq,\bmu)|\tilde{a}_{x,\sigma}(t)|\boldsymbol{\psi}_{N+1,M}(\boldsymbol{k},
\bl)\rangle\, ,
\ee
is a general form factor of the $\tilde{a}_{x,\sigma}(t)$ operator (the form factor of $\tilde{a}^\dagger_{x,\sigma}(t)$ is obtained by complex conjugation) and
\be
\bar{M}=\left\{\begin{array}{lll} M&\ \ \mbox{ if } &\sigma=\uparrow\, ,\\
M-1&\ \ \mbox{ if } &\sigma=\downarrow\, .
\end{array}\right.
\ee
The determinant representations are obtained by following these steps \cite{IPA98, IP98}: (a) computing the form factors for a finite-size system; (b) deriving 
the determinant formula for the expectation values of bilocal operators using a technique similar to the Cauchy-Binet formula; and (c) taking the thermodynamic 
limit by applying von Koch's determinant formula.

\end{widetext}

\end{document}